# A Systematic Framework to Identify Hidden Vulnerabilities in Three-Fold Redundant Sparse Arrays Under Dual Sensor Failures

Ashish Patwari*, Budhaditya Bhattacharyya, and Lavanya Nagarajan

***Abstract*—As the field of sparse arrays progressed, numerous array designs have been introduced with a focus on larger apertures and higher degrees of freedom (DOFs), resulting in maximally economic sparse arrays (MESAs) that operate with the least number of sensors required to provide a given aperture while ensuring a hole-free difference coarray (DCA). Consequently, MESAs are least robust to sensor failures and cannot afford the failure of even a single sensor. Multifold redundant sparse arrays (MFRSAs) provide a practical solution to the problem of sensor failures in sparse arrays by making sure that the array contains enough sensor pairs necessary to produce each spatial lag multiple times. Owing to this property, a β-fold redundant array can withstand simultaneous failure of at least β-1 sensors without losing the hole-free DCA property. Nevertheless, MFRSAs are also prone to hidden dependencies that prevent them from being fully robust. In this work, we present a systematic framework to evaluate the robustness of triple redundant sparse linear arrays (TRSLAs) against all possible two-sensor failures. After detailing the proposed approach, we present the failure analysis of representative TRSLAs available in existing literature. It is found that existing TRSLAs have some hidden vulnerabilities against the failure of some peculiar sensor pairs. Corresponding MATLAB programs and numerical simulations are provided for evaluation and use by the array processing community. The proposed approach has a great archival value as it can evaluate the robustness of any present or future TRSLAs through objective means.**

## I. INTRODUCTION

Sensor arrays find applications in various fields such as radar, sonar, wireless communications, seismic signal processing, medical imaging, sound source localization etc [1], [2]. A sensor array consists of two or more sensors arranged in a specific geometry. Sensors such as antennas, microphones, hydrophones, geophones etc., are used in the above applications to determine the directions of the incoming wave fields that impinge the array [3]. While uniform arrays have been used traditionally, modern applications are exploring sparse arrays. Sparse arrays can provide the same aperture (angular resolution) as that of uniform arrays using fewer sensors and hence offer huge savings in system cost, power requirements, installation and maintenance, heating and cooling etc. Additionally, they are less affected by mutual coupling and can detect more source directions than full arrays [4]. Owing to these advantages, sparse arrays offer tremendous possibilities in future systems such as Terahertz communication (THz), integrated sensing and communications (ISAC), reconfigurable intelligent surfaces (RIS), autonomous driving, audio enhancement etc. [5].

The field of sparse array design has prospered beyond imagination in the past fifteen years, following the introduction of co-prime and nested arrays [6]. Sparse array analysis is generally carried out in the coarray domain using the concept of second order difference coarray (DCA). Most of the existing sparse array designs focus on providing the largest aperture, the largest hole-free coarray, and the least mutual coupling etc., for a given number of sensors [7], [8], [9], [10], [11]. Very few designs focus on robustness to sensor failures [12], [13], [14], [15], [16], [17]. An otherwise hole-free array might also encounter holes in the DCA during sensor failures. Presence of holes in the DCA leads to ambiguities in the DOA estimation process as explained below [18], [19].

Three options are available to overcome the effect of holes in the DCA. First is the use of compressed sensing methods for DOA estimation which do not depend on the continuity of the coarray [20], [21]. Second is to employ techniques such as coarray interpolation/hole-filling/aperture extension/matrix completion etc., to work around the discontinuities in the coarray [22], [23], [24]. The third and the most useful method, especially when coarray processing is employed for DOA estimation is the use of multi-fold redundant sparse array (MFRSA) configurations that have inbuilt robustness against sensor failures.

A handful of two-fold and three-fold linear arrays have been proposed in existing literature [13], [14], [16], [25], [26]. Such arrays must satisfy stringent mathematical requirements, failing which their usability becomes questionable under certain

Corresponding author: Ashish Patwari (ashish.p@vit.ac.in)
All authors are affiliated with the School of Electronics Engineering, Vellore Institute of Technology, Vellore, Tamil Nadu, India 632 014 (email: ashish.p@vit.ac.in; budhaditya@vit.ac.in ; lavanya.n@vit.ac.in ) circumstances. To ensure that the candidate arrays behave properly, thorough checking of their properties is necessary. This forms the basis for the rest of the discussion in this paper.

We propose a systematic framework to evaluate the robustness of triple redundant SLAs (TRSLAs). Using this framework, one can instantly check whether a given array abides by the three-fold redundancy requirements or has any shortcomings that prevent it from doing so. *The main contributions of this work are:*

- A systematic framework has been developed to verify the robustness of TRSLAs against all permissible two sensor failures. A MATLAB program for the same is also provided. To the best of our knowledge, there are no readymade tools in the existing literature for the same.
- The term hidden essential sensor pairs (HESPs) is coined for the first time to aid in the analysis of sparse arrays. Thorough failure analysis of the 3FRA and the TRA revealed the presence of HESPs whose failure can disrupt the DCA continuity.
- A closed-form expression governing the relation between the array aperture, position of the HESPs, and hole location has been established. Based on these, we propose modified versions of the 3FRA and the TRA which are devoid of any HESPs.

The rest of the paper is structured as follows. Section II provides relevant sparse array terminology. Section III presents the proposed methodology to detect the robustness of TRSLAs. Section IV describes the methodology used to obtain simulation results. Section V provides numerical simulation results. Discussion is provided in section VI, and section VII concludes the paper. MATLAB routines are given in the appendix section.

Note: In this paper, the term TRSLA represents the whole family of sparse arrays that follow the three-fold redundancy requirements and robustness to two-sensor failures. The term FRA represents the array proposed by Dong et al. and the TRA refers to array proposed by Hou et al.

## II. Related Background

This section explains the concepts related to sparse arrays in general and TRSLAs in particular.

### *A. Sparse Array Terminology*

#### *1) Difference Coarray, Holes, and Degrees of Freedom*

Let $\mathbb{S}$ denote the position set of physical sensors in the array, relative to half wavelength. The difference set $\mathbb{Z}$ is obtained by evaluating all possible self- and cross- subtractions between the physical sensor positions. Each entry in $\mathbb{Z}$ denotes a spatial lag. The difference coarray (DCA) $\mathbb{D}$ contains the sorted and non-repeating elements of $\mathbb{Z}$. Missing spatial lags in $\mathbb{Z}$ or $\mathbb{D}$ are called holes.

The central hole-free portion of the DCA is given by $\mathbb{U}$ and denotes the uniform degrees of freedom (uDOFs) offered by the array. DOFs represent the number of resolvable source angles that the array can estimate during DOA estimation. For arrays whose DCAs are hole-free, it follows $\mathbb{D} = \mathbb{U}$.

#### *2) Weight Function*

The weight function $w(m)$ denotes the number of element pairs separated by distance *m*. Each entry in the weight function signifies the number of times a given spatial lag appears in the difference set $\mathbb{Z}$.

#### *3) Desirable features of TRSLAs (Weight Distribution and Essential Sensor Pairs)*

The desirable properties of β-fold redundant arrays (βFRAs) have been clearly outlined in [16] and the concepts of k-essentialness were covered in [18]. By combining both these analyses, the following requirements can be specified for any TRSLA.

*a) Weight Function Distribution*

If $L$ denotes the array aperture and $[-L_u, L_u]$ denotes the central portion of the DCA wherein all spatial lags have a weight of three or more, it follows that $L_u = L - 2$. The weights of the last two spatial lags shall follow the relation $w(m) = L - |m| + 1$. Accordingly, the weight function of a TRSLA can be represented in a compact form by

$$w(i) \geq 3;\ 0 \leq i \leq L_u \quad w(L-1) = 2 \quad w(L) = 1 \tag{1}$$

The negative portion of the weight function is automatically defined due to the even symmetry, $w(-m) = w(m)$.

*b) Essential Sensor Pairs*

A TRSLA shall be robust to all two-element failures (barring the failure of essential sensor pairs). Two sensors constitute an essential sensor pair (ESP) if their combined removal from the physical array either alters the span or the continuity of the DCA [18]. The following are the valid ESPs for any TRSLA

$$\mathcal{E}_2 = \{(0,n),(m,L),(1,L-1)\};$$
$$1 \leq n \leq L\,, 1 \leq m \leq L-1 \tag{2}$$

The cardinality of $\mathcal{E}_2$ denotes the number of ESPs and is given by $|\mathcal{E}_2| = 2N - 2$ [19]. Therefore, a valid TRSLA consists of exactly $2N - 2$ ESPs. Additionally, ESPs occurring at positions other than those mentioned in the set $\mathcal{E}_2$ are labeled as hidden essential sensor pairs (HESPs). HESPs are undesirable as their presence can cause ambiguities in the DOA estimation process, especially when coarray processing is employed.

*4) Inter-element spacing (IES) notation*

The IES notation lists the separations between successive elements in the physical array. Hence, the IES form of a $N$-element array has $N-1$ entries. An array with the IES notation $\{a, b, c, d\}$ has physical sensors at $\{0, a, a+b, a+b+c, a+b+c+d\}$, relative to half wavelength. This notation has been widely utilized in the past for sparse array representation [27], [28], [29], [30] and has come back to the limelight with the introduction of the maximum inter-element spacing criterion (MISC) array [9].

*B. Coarray MUSIC*

The coarray MUSIC algorithm is widely used for DOA estimation in sparse arrays. The signal model for second order difference coarray processing is based on the Eigen value decomposition of the coarray correlation matrix and is widely available in the existing literature.

## III. Proposed Method

A systematic testing methodology is developed to check the robustness of the TRA to against all permissible two-element failures. Initially, manual methods were employed. Two sensors were removed at a given time from the TRA, and the weight function of the resultant TRA was computed. The same was repeated for all sensor pairs. However, the process was very tedious because even a small array with $N = 15$ consists of $15_{C_2} = 105$ sensor pairs. Barring the $2N - 2 = 28$ essential sensor pairs, there would still be 77 combinations of two-sensor failures against which the array must be tested. Clearly, this process of manual testing was cumbersome and intractable with growing $N$. To address this, we developed a MATLAB program that can automatically create all permissible two-element failures and test the array's weight function/DCA continuity in each case. The program returns two outputs as shown in Fig. 1. If the array under test (AUT) passes all checks, it displays the message "Array provides triple redundancy" in the MATLAB command window. Else, the positions of HESPs are returned along with the message "Array cannot provide triple redundancy". Figure 2 demonstrates the algorithmic approach for HESP detection in detail.

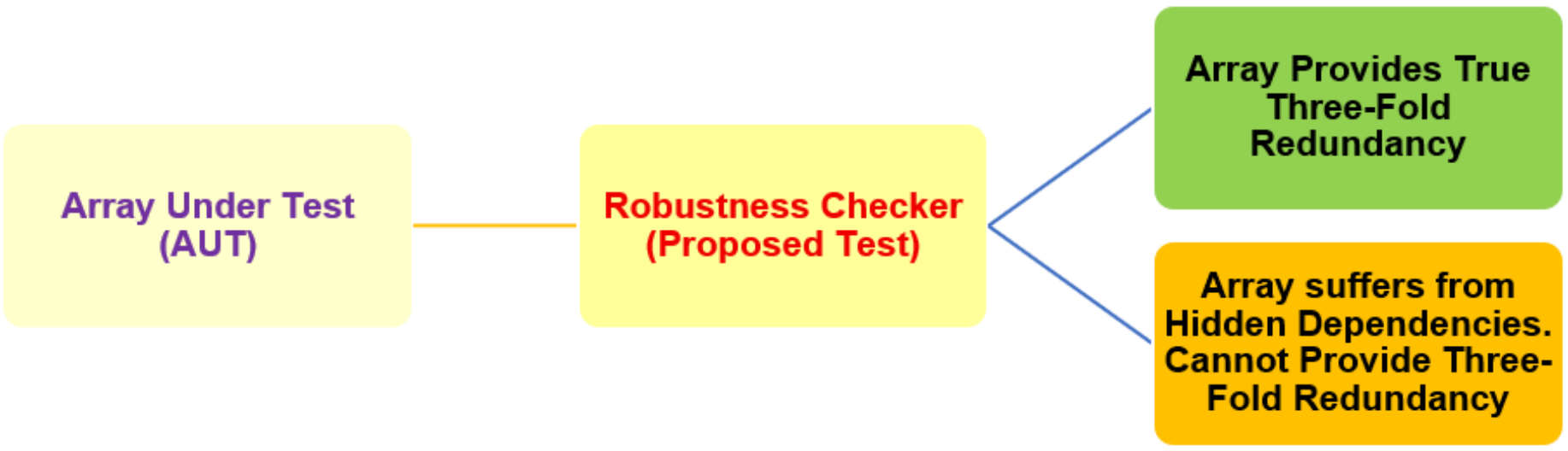

Fig. 1. Proposed Test for Verifying Three-Fold Redundancy

Exhaustive simulations have been carried out by us to verify the correctness of the proposed approach and program, as described next.

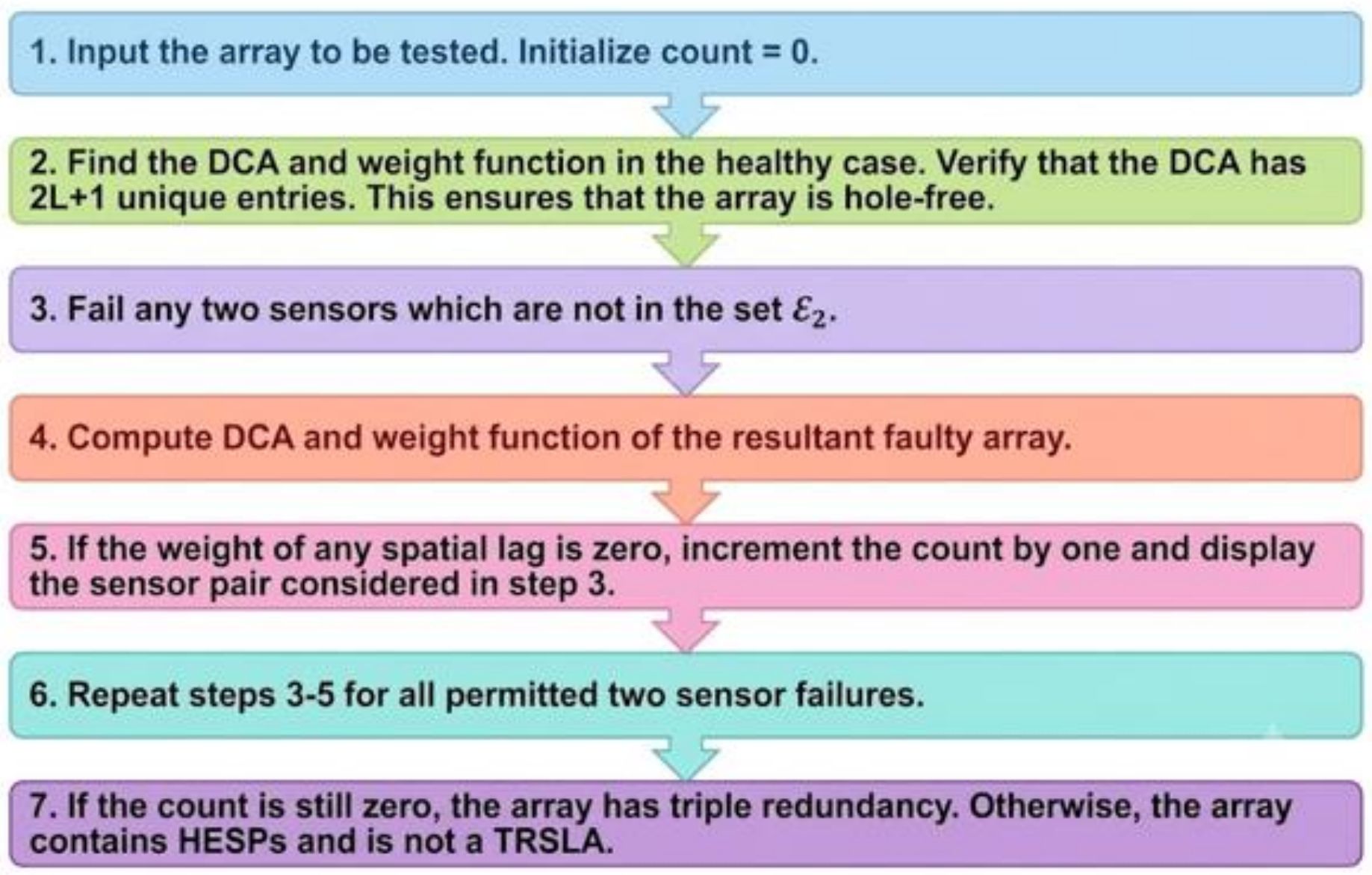

Fig. 2. Methodology to perform failure analysis of three-fold redundant arrays

## IV. Simulation Methodology

We followed the following methodology to carry out simulations. First, MATLAB programs were written to realize (i) Dong's 3FRA for any size above N=9 and (ii) TRA for any size above N=15. After verifying the correctness of the programs, several arrays of varying sizes (number of sensors) have been obtained and passed through the testing program to diagnose whether they are truly robust to all two-sensor failures or contain some HESPs. For each array size (say, N=12), the array aperture, the positions of HESPs (if any) have been noted.

Finally, the correctness of the proposed framework/program has been reinforced by plotting the weight function of candidate arrays in three cases: - healthy case, faulty case where HESPs are failed, faulty case where sensors at other positions than indicated by HESPs are failed. It is verified that while general instances of two sensor failures do not affect the continuity of DCA, failure of HESPs **indeed** form holes in the DCA and hence cause gaps in the weight function graph. This verifies the correctness of the proposed program.

## V. Numerical Results

The following results were obtained when the proposed approach was used to test the robustness of existing TRSLAs.

### *A. Correctness of the Proposed Method*

To steer clear of any doubts about the correctness of the proposed method, we show the weight function analysis of a representative array under different scenarios.

For instance, consider the 15-element 3FRA. As per definition, its sensors are positioned at [0, 1, 2, 6, 7, 8, 15, 16, 17, 24, 25, 26, 27, 28, 29]. In the healthy case, its weight function is as shown in Fig. 3. The weight function follows the expected distribution as outlined in (1). Upon passing the above array as an input to the proposed test program, it revealed HESPs at (2, 15) and (15, 16), respectively.

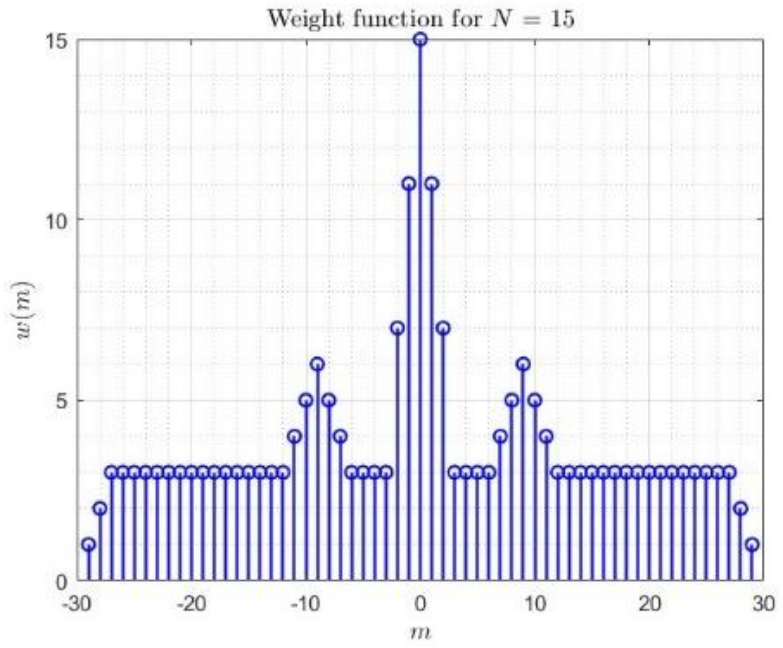

Fig. 3. Weight function of the 3FRA in healthy case (N=15)

Next, we show the weight function of the above 3FRA when two of its sensors fail. Firstly, two sensors that do not belong to the HESPs are failed. Figure 4 shows the weight function of the faulty 3FRA when the sensors at (6, 8) fail. As seen in Fig. 4, all spatial lags occur at least once, thereby representing a hole-free DCA, as desired.

On the other hand, the situation would be different if the failed sensors belonged to the HESP list. Figure 5 shows the weight function when the HESP (2, 15) is failed. It can be observed that spatial lag 14 is missing. This shrinks the central continuous portion of the DCA to [-13, +13] from [-29, +29], thereby reducing the uDOFs from 59 to 27. Although signal processing techniques such as coarray interpolation, aperture extension, matrix completion etc., provide a way around to work with the discontinuities in the DCA, they are computationally expensive. Moreover, the occurrence of holes in the DCA during a two-sensor failure is against the very notion of providing three-fold redundancy in the first place.

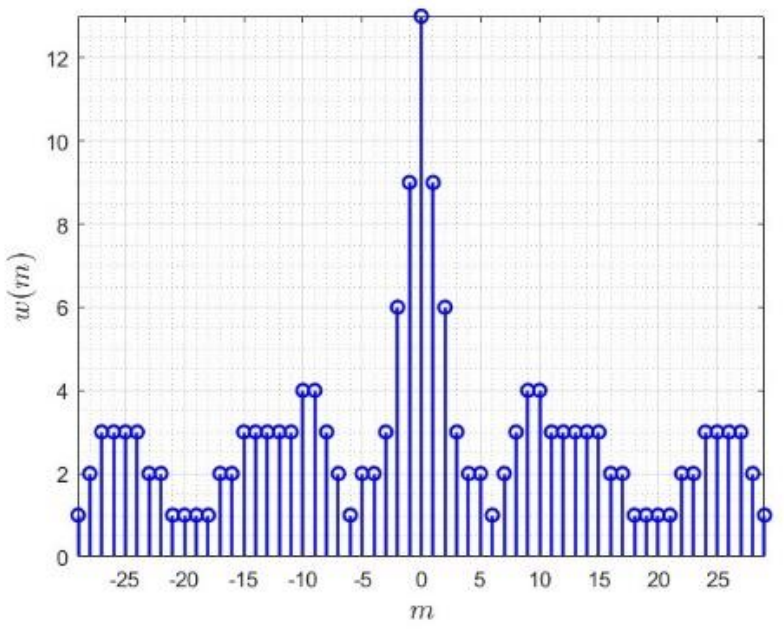

Fig. 4. 15-element 3FRA with sensor failures at 6 and 8

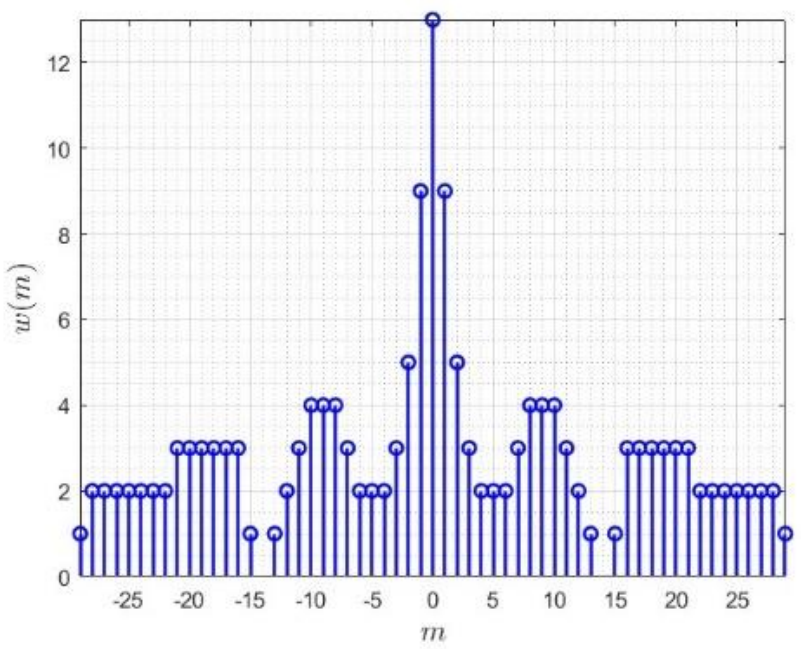

Fig. 5. Missing spatial lags when HESPs (2, 15) fail

A similar analysis was repeated on 3FRAs and TRAs of various sizes to check for the presence of HESPs. The correctness of the proposed test program and its ability to detect hidden dependencies (or the lack of them) in a given TRSLA configuration have been verified. It is observed that the failure of HESPs *actually* disrupt the DCA.

### *B. Failure analysis of Dong's 3FRA*

Firstly, 3FRAs for various array sizes ($N$ = 10 to 50) were generated using the program given in appendix B. The generated arrays were passed as an input to the test program described in the previous section. The test revealed that some 3FRAs suffer from the presence of HESPs as shown in Table I.

Table I – Array Configurations and Hidden Dependencies for various 3FRAs

| N | Array Configuration | L | HESP locations |
|---|---|---|---|
| 10 | 0 1 2 4 5 6 11 12 13 14 | 14 | [5,6] and [6,11] |
| 11 | 0 1 2 5 6 7 13 14 15 16 17 | 17 | [6,7] and [7,13] |
| 12 | 0 1 2 6 7 8 15 16 17 18 19 20 | 20 | [7,8] and [8,15] |
| 13 | 0 1 2 7 8 9 17 18 19 20 21 22 23 | 23 | [8,9] and [9,17] |
| 14 | 0 1 2 8 9 10 19 20 21 22 23 24 25 26 | 26 | [9,10] and [10,19] |
| 15 | 0 1 2 6 7 8 15 16 17 24 25 26 27 28 29 | 29 | [2,15] and [15,16] |
| 16 | 0 1 2 7 8 9 17 18 19 27 28 29 30 31 32 33 | 33 | [2,17] and [17,18] |
| 17 | 0 1 2 8 9 10 19 20 21 30 31 32 33 34 35 36 37 | 37 | [2,19] and [19,20] |
| 18 | 0 1 2 9 10 11 21 22 23 33 34 35 36 37 38 39 40 41 | 41 | [2,21] and [21,22] |
| 19 | 0 1 2 10 11 12 23 24 25 36 37 38 39 40 41 42 43 44 45 | 45 | [2,23] and [23,24] |
| 20 | 0 1 2 11 12 13 25 26 27 39 40 41 42 43 44 45 46 47 48 49 | 49 | [2,25] and [25,26] |
| 21 | 0 1 2 9 10 11 21 22 23 33 34 35 45 46 47 48 49 50 51 52 53 | 53 | Nil |
| 22 | 0 1 2 10 11 12 23 24 25 36 37 38 49 50 51 52 53 54 55 56 57 58 | 58 | Nil |
| 23 | 0 1 2 11 12 13 25 26 27 39 40 41 53 54 55 56 57 58 59 60 61 62 63 | 63 | Nil |
| 24 | 0 1 2 12 13 14 27 28 29 42 43 44 57 58 59 60 61 62 63 64 65 66 67 68 | 68 | Nil |
| 25 | 0 1 2 13 14 15 29 30 31 45 46 47 61 62 63 64 65 66 67 68 69 70 71 72 73 | 73 | Nil |
| 26 | 0 1 2 14 15 16 31 32 33 48 49 50 65 66 67 68 69 70 71 72 73 74 75 76 77 78 | 78 | Nil |
| 27 | 0 1 2 12 13 14 27 28 29 42 43 44 57 58 59 72 73 74 75 76 77 78 79 80 81 82 83 | 83 | [2,42] and [42,43] |
| 28 | 0 1 2 13 14 15 29 30 31 45 46 47 61 62 63 77 78 79 80 81 82 83 84 85 86 87 88 89 | 89 | [2,45] and [45,46] |
| 29 | 0 1 2 14 15 16 31 32 33 48 49 50 65 66 67 82 83 84 85 86 87 88 89 90 91 92 93 94 95 | 95 | [2,48] and [48,49] |
| 30 | 0 1 2 15 16 17 33 34 35 51 52 53 69 70 71 87 88 89 90 91 92 93 94 95 96 97 98 99 100 101 | 101 | [2,51] and [51,52] |
| 31 | 0 1 2 16 17 18 35 36 37 54 55 56 73 74 75 92 93 94 95 96 97 98 99 100 101 102 103 104 105 106 107 | 107 | [2,54] and [54,55] |
| 32 | 0 1 2 17 18 19 37 38 39 57 58 59 77 78 79 97 98 99 100 101 102 103 104 105 106 107 108 109 110 111 112 113 | 113 | [2,57] and [57,58] |
| 33 | 0 1 2 15 16 17 33 34 35 51 52 53 69 70 71 87 88 89 105 106 107 108 109 110 111 112 113 114 115 116 117 118 119 | 119 | Nil |
| 34 | 0 1 2 16 17 18 35 36 37 54 55 56 73 74 75 92 93 94 111 112 113 114 115 116 117 118 119 120 121 122 123 124 125 126 | 126 | Nil |
| 35 | 0 1 2 17 18 19 37 38 39 57 58 59 77 78 79 97 98 99 117 118 119 120 121 122 123 124 125 126 127 128 129 130 131 132 133 | 133 | Nil |
| 36 | 0 1 2 18 19 20 39 40 41 60 61 62 81 82 83 102 103 104 123 124 125 126 127 128 129 130 131 132 133 134 135 136 137 138 139 140 | 140 | Nil |
| 37 | 0 1 2 19 20 21 41 42 43 63 64 65 85 86 87 107 108 109 129 130 131 132 133 134 135 136 137 138 139 140 141 142 143 144 145 146 147 | 147 | Nil |
| 38 | 0 1 2 20 21 22 43 44 45 66 67 68 89 90 91 112 113 114 135 136 137 138 139 140 141 142 143 144 145 146 147 148 149 150 151 152 153 154 | 154 | Nil |
| 39 | 0 1 2 18 19 20 39 40 41 60 61 62 81 82 83 102 103 104 123 124 125 144 145 146 147 148 149 150 151 152 153 154 155 156 157 158 159 160 161 | 161 | [2,81] and [81,82] |
| 40 | 0 1 2 19 20 21 41 42 43 63 64 65 85 86 87 107 108 109 129 130 131 151 152 153 154 155 156 157 158 159 160 161 162 163 164 165 166 167 168 169 | 169 | [2,85] and [85,86] |

In fact, we tested 3FRA configurations until N=100 but have not shown here due to space constraints. A regular pattern has been observed regarding the vulnerability of 3FRAs to two-sensor failures. Six consecutive arrays, starting at N=15, contain HESPs. The next six are robust (devoid of HESPs). The same cycle repeats periodicity six. The vulnerable array configurations contain HESPs at $\left(2,\frac{L+1}{2}\right)$ and $\left(\frac{L+1}{2},\frac{L+3}{2}\right)$, where L is the array aperture. Failure of either pair creates a hole at $\pm\left(\frac{L-1}{2}\right)$ in DCA. This factor cannot be neglected as this effect is not limited to a particular array size. As shown in Table I, almost 50% of the 3FRA configurations suffer from this problem.

The findings from Table I can be summarized using the following expressions. Define two scalars $k$ and $i$ defined by $k = mod(N-15,6)$ and $i = \frac{N-15-k}{6}$. The robustness or vulnerability of a given 3FRA depends on the value of the scalar $i$. If $i$ is even, the array contains HESPs. Otherwise, it is truly robust against two-sensor failures. This formulation can be verified with a numerical example. Consider $N = 30$. This implies $k = 3$ and $i = 2$, as per their respective definitions. Since $i$ is even, it can be concluded that the given 3FRA has HESPs. This is consistent with the results depicted in Table I.

### *C. Failure analysis of the Ternary Redundant Array*

A similar exercise was repeated for TRAs. TRAs of various sizes were thoroughly analyzed for robustness using the test program described in the previous section. For each array size, the HESPs (if any), have been noted. Table II shows the HESP locations in TRAs from $N = 15$ to 30.

TABLE II Hidden Essential Sensor Pairs In The TRA

| *N* | Array Configurations | *L* | HESP locations |
|---|---|---|---|
| 15 | 0 1 2 3 4 5 8 11 14 17 20 23 25 27 29 | 29 | [2, 27], [3, 27], [4, 14], [11, 14] & [14, 17] |
| 16 | 0 1 2 3 4 5 8 11 14 17 20 23 25 27 29 31 | 31 | [2, 29], [3, 29], [4, 14] & [14, 17] |
| 17 | 0 1 2 3 4 5 6 9 12 15 18 21 24 27 29 31 33 | 33 | [2, 31], [3, 31], [4, 15] & [15, 18] |
| 18 | 0 1 2 3 4 5 6 9 12 15 18 21 24 27 29 31 33 35 | 35 | [2, 33], [3, 33], [2, 18] & [15, 18] |
| 19 | 0 1 2 3 4 5 6 7 10 13 16 19 22 25 28 31 33 35 37 | 37 | [2, 35], [3, 35], [2, 19] & [16, 19] |
| 20 | 0 1 2 3 4 5 6 7 10 13 16 19 22 25 28 31 33 35 37 39 | 39 | [2, 37] & [3, 37] |
| 21 | 0 1 2 3 4 5 6 7 8 11 14 17 20 23 26 29 32 35 37 39 41 | 41 | [2, 39] & [3, 39] |
| 22 | 0 1 2 3 4 5 6 7 8 11 14 17 20 23 26 29 32 35 37 39 41 43 | 43 | [2, 41] & [3, 41] |
| 23 | 0 1 2 3 4 5 6 7 8 9 12 15 18 21 24 27 30 33 36 39 41 43 45 | 45 | [2, 43] & [3, 43] |
| 24 | 0 1 2 3 4 5 6 7 8 9 12 15 18 21 24 27 30 33 36 39 41 43 45 47 | 47 | [2, 45] & [3, 45] |
| 25 | 0 1 2 3 4 5 6 7 8 9 10 13 16 19 22 25 28 31 34 37 40 43 45 47 49 | 49 | [2, 47] & [3, 47] |
| 26 | 0 1 2 3 4 5 6 7 8 9 10 13 16 19 22 25 28 31 34 37 40 43 45 47 49 51 | 51 | [2, 49] & [3, 49] |
| 27 | 0 1 2 3 4 5 6 7 8 9 10 11 14 17 20 23 26 29 32 35 38 41 44 47 49 51 53 | 53 | [2, 51] & [3, 51] |
| 28 | 0 1 2 3 4 5 6 7 8 9 10 11 14 17 20 23 26 29 32 35 38 41 44 47 49 51 53 55 | 55 | [2, 53] & [3, 53] |
| 29 | 0 1 2 3 4 5 6 7 8 9 10 11 12 15 18 21 24 27 30 33 36 39 42 45 48 51 53 55 57 | 57 | [2, 55] & [3, 55] |
| 30 | 0 1 2 3 4 5 6 7 8 9 10 11 12 15 18 21 24 27 30 33 36 39 42 45 48 51 53 55 57 59 | 59 | [2, 57] & [3, 57] |
| 31 | 0 1 2 3 4 5 6 7 8 9 10 11 12 13 16 19 22 25 28 31 34 37 40 43 46 49 52 55 57 59 61 | 61 | [2, 59] & [3, 59] |
| 32 | 0 1 2 3 4 5 6 7 8 9 10 11 12 13 16 19 22 25 28 31 34 37 40 43 46 49 52 55 57 59 61 63 | 63 | [2, 61] & [3, 61] |
| 33 | 0 1 2 3 4 5 6 7 8 9 10 11 12 13 14 17 20 23 26 29 32 35 38 41 44 47 50 53 56 59 61 63 65 | 65 | [2, 63] & [3, 63] |
| 34 | 0 1 2 3 4 5 6 7 8 9 10 11 12 13 14 17 20 23 26 29 32 35 38 41 44 47 50 53 56 59 61 63 65 67 | 67 | [2, 65] & [3, 65] |
| 35 | 0 1 2 3 4 5 6 7 8 9 10 11 12 13 14 15 18 21 24 27 30 33 36 39 42 45 48 51 54 57 60 63 65 67 69 | 69 | [2, 67] & [3, 67] |
| 36 | 0 1 2 3 4 5 6 7 8 9 10 11 12 13 14 15 18 21 24 27 30 33 36 39 42 45 48 51 54 57 60 63 65 67 69 71 | 71 | [2, 69] & [3, 69] |
| 37 | 0 1 2 3 4 5 6 7 8 9 10 11 12 13 14 15 16 19 22 25 28 31 34 37 40 43 46 49 52 55 58 61 64 67 69 71 73 | 73 | [2, 71] & [3, 71] |
| 38 | 0 1 2 3 4 5 6 7 8 9 10 11 12 13 14 15 16 19 22 25 28 31 34 37 40 43 46 49 52 55 58 61 64 67 69 71 73 75 | 75 | [2, 73] & [3, 73] |
| 39 | 0 1 2 3 4 5 6 7 8 9 10 11 12 13 14 15 16 17 20 23 26 29 32 35 38 41 44 47 50 53 56 59 62 65 68 71 73 75 77 | 77 | [2, 75] & [3, 75] |
| 40 | 0 1 2 3 4 5 6 7 8 9 10 11 12 13 14 15 16 17 20 23 26 29 32 35 38 41 44 47 50 53 56 59 62 65 68 71 73 75 77 79 | 79 | [2, 77] & [3, 77] |

The same exercise has been repeated till N=50 and beyond, only to find a similar pattern. It can be inferred from Table II that all TRAs contain two or more HESPs. To be specific, all TRAs with $N \geq 20$ contain exactly two HESPs, located at $(2, L-2)$ and $(3, L-2)$, respectively. The failure of any one sensor pair among the two creates a hole at $\pm(L-1)$ in the DCA. This is mainly because of improper terminal weights in the TRA. Hence, it can be concluded that both 3FRA and TRA suffer from HESPs.

## VI. Discussion

As happens with most complex phenomena, finding an optimum TRSLA configuration is a difficult (NP-hard) computational task, but verifying whether a given array satisfies the three-fold requirement is not computationally so expensive. That said, the verification process is not so simple either because it involves a huge number of computations and combinations to be checked.

For example, in an array with 310 sensors, there would be 47,895 sensor pairs. Barring the 618 essential sensor pairs, there would be 47,277 cases of two sensor failures against which the array has to be checked. This computation took around 4 minutes on an Intel core i7 – 7500U processor with Windows 11 operating system, 2.7 GHz clock, and 16 GB RAM using MATLAB. However, arrays with N < 200 could be verified in less than one minute.

The aforementioned execution times also prove the effectiveness of the proposed approach. It would be impossible to manually create 47,277 failure scenarios and compute the weight function in each case. It is fairly beyond human capacity.

It is important to ensure that MFRAs do not have any extra/hidden essential sensors or essential sensor pairs, apart from those at positions already agreed upon. For example, as per definition, a two-fold redundant array shall have only two essential sensors, namely, at 0 and L. As the array designer already knows the importance of these two sensors in preserving the array aperture (or DCA span), he/she would take enough care to ensure that these sensors do not fail. But if the array has HESPs, the array designer would be unaware of it. This could be fatal when such arrays are deployed in practical systems. The same argument is valid for three-fold and four-fold redundant sparse arrays that have HESPs and sensor triads, respectively.

## VII. Conclusion

A systematic framework and a MATLAB program to verify the resilience of triple redundant sparse linear arrays (TRSLAs) has been proposed. The correctness of the proposed approach has been verified by analyzing the changes in the weight function under healthy and faulty conditions.

It is found that existing TRSLAs suffer from the presence of HESPs. Nearly 50% of 3FRAs and all TRAs contain HESPs. 3FRAs have HESPs located at $\left(2,\frac{L+1}{2}\right)$ and $\left(\frac{L+1}{2},\frac{L+3}{2}\right)$, while TRAs have them at $(2, L-2)$ and $(3, L-2)$, respectively. Failure of HESPs creates holes in the DCA, which is against the very principle of providing three-fold redundancy.

Similar mechanisms could be developed in the future to check the robustness four-fold arrays against triple sensor failures, respectively. It is worth mentioning that the proposed approach is a universal tool to check the robustness of any present or future TRSLAs.

## APPENDIX A

### A.1 3FRA formulation

The formulation of Zhu's β-FRA is given in equations (31) and (32) of [16]. A 3FRA is a special case of the β-FRA. A MATLAB program is also provided to generate any desired 3FRA configuration (for $N \geq 9$). The IES notation of the β-FRA is given by the closed-form expression

$$\mathbb{I}_{\beta FRA^*} = \{1^{\beta-1},\ p^*,\ (1^{\beta-1},\ p^*+\beta)^{\ m^*},\ 1^{p^*+\beta-2}\} \tag{A.1}$$

where the optimal values of parameters $p,\ m$, for maximizing $L_u\ (= L-2)$ are given by

$$m^* = \left\lfloor \frac{N}{2\beta} \right\rceil - 1$$
$$p^* = N - (m^*+2)\beta + 1 \tag{A.2}$$

The symbol $\lfloor\ \bullet\ \rceil$denotes rounding to the nearest integer. The 3FRA can be obtained by letting $\beta$=3 in the above formulations. Care has to be taken to convert these IES notations into the actual sensor positions first before beginning any analysis.

### A.2 TRA formulation

The ternary redundant array (TRA) was proposed to provide three-fold redundancy but with lower susceptibility to mutual coupling than Dong's 3FRA. As the original paper describing the TRA is in Chinese language [26], a quick review of its formulation and array geometry is presented below.

The TRA requires 15 or more available sensors ($N \geq 15$). The IES notation of the the TRA is given by

$$\mathbb{I}_{TRA} = \{1^{N_1}, 3^{N_2}, 2^{N_3-1}\}$$
$$s.t.$$
$$N_2 = \left\lfloor \frac{N+1}{2} \right\rceil - 2;\ \ N_1 = N_2 - 1;\ \ N_3 = N - N_1 - N_2 \tag{A.3}$$

The sensors in the array are arranged into three subarrays, each consisting of $N_1+1,\ N_2$, and $N_3-1$ sensors, respectively, such that $N = N_1+N_2+N_3$. The first subarray has $N_1+1$ sensors with unit spacing, the second has $N_2$ sensors with triple spacing, and the third has $N_3-1$ sensors with double spacing, leading to the sensor positions $\mathbb{S}_1 = \{0,1,2,\dots,\ N_1\}$, $\mathbb{S}_2 = \{N_1+3,\ N_1+6,\dots,\ N_1+3N_2\}$ and $\mathbb{S}_3 = \{N_1+3N_2+2, N_1+3N_2+4,\dots,2N-1\}$, respectively.

The complete array is therefore given by $\mathbb{S} = \mathbb{S}_1 \cup \mathbb{S}_2 \cup \mathbb{S}_3$. Accordingly, the first three and the last three sensors of the TRA would be $[0, 1, 2, \dots, 2N-5, 2N-3, 2N-1]$, thus providing an aperture of $L = 2N-1$ and $D = 4N-1$ uniform DOFs. A MATLAB program that can generate TRAs of any specified array size ($N \geq 15$) is provided below.